\magnification=\magstep1
\baselineskip=20 truept
\def\ni{\noindent}

\def\bn{\bigbreak\noindent}
\def\sn{\smallskip\noindent}
\def\ie{{\it i.e.}}
\def\r2{R^2}
\def\pa{\partial}
\def\half{{\textstyle{1\over2}}}
\def\quar{{\textstyle{1\over4}}}
\def\vp{\varphi}
\def\ve{\varepsilon}
\font\scap=cmcsc10
\font\title=cmbx10 scaled\magstep1
\rightline{DTP96/5}
\vskip 1truein
\centerline{\title Bogomol'nyi bounds for two-dimensional}
\centerline{\title lattice systems.}

\vskip 1truein
\centerline{\scap R. S. Ward}

\bn\centerline{\it Dept of Mathematical Sciences, University of Durham,}
\centerline{\it Durham DH1 3LE, UK.}

\vskip 2truein
\ni{\bf Abstract.} The O(3) sigma model and abelian Higgs model in two
space dimensions admit topological (Bogomol'nyi) lower bounds on their energy.
This paper proposes lattice versions of these systems which maintain the
Bogomol'nyi bounds. One consequence is that instantons/solitons/vortices
on the lattice then have a high degree of stability.
\vskip 1truecm
\ni To appear in Communications in Mathematical Physics.
\vfil\eject

\ni{\bf1. Introduction}

\sn In systems where there are topological configurations (instantons,
vortices, Skyrmions, monopoles, etc) classified by an integer $k$, one often
has a Bogomol'nyi bound $E\geq\alpha|k|$, where $E$ is the energy (or action)
of the system, and $\alpha$ is a universal constant. The stability of the
topological objects in question is often related to the existence of such a
lower bound on $E$.
Lattice versions of these systems have been much studied (for
purposes of numerical computation, or regularization of the quantum field
theory, etc), but the Bogomol'nyi bound has generally not been preserved on
the lattice. Although one can identify topological objects in the
lattice systems, and the details of how to do this are well-known, these
topological objects are often unstable.

Given a continuum field theory, there are many different lattice systems
which reduce to it in the continuum limit. The object of this paper is to
present lattice versions of two systems in which the Bogomol'nyi bound
is maintained. The systems in question are the O(3) sigma
model and the abelian Higgs model, both in two space dimensions. Because
the Bogomol'nyi bound is satisfied, topological objects will be well-behaved
even when their size is not much greater than the lattice spacing.

A similar study was undertaken a few years ago for systems in one space
dimension [16, 17]. This concentrated on the sine-Gordon system, but analogous
results hold for other systems (such as phi-fourth) which admit kink solutions.
In these one-dimensional cases, the Bogomol'nyi bound can be attained: there
exists a one-real-parameter family of static kink solutions on the lattice,
with energy equal to the topological minimum. The parameter describes the
position of the kink on the one-dimensional lattice: in particular, its energy
is the same irrespective of where it is in relation to the lattice (there is
no Peierls-Nabarro energy barrier). These
features are not present in the two-dimensional cases discussed in this paper;
the energy of a topological object will be minimized only when it is located at
the centre of a plaquette, and even then is greater than the topological
minimum.

If one imposes radial symmetry in the plane, then in effect one obtains a
one-dimensional system, and one can try to set up a lattice version of this
which maintains the Bogomol'nyi bound. This was done for the O(3)
sigma-model [5], and used to study the stability of single
(radially-symmetric) solitons [7]. The case of semilocal vortices in an
abelian Higgs model (with Higgs doublet) was dealt with in the same way [6].
But the intention in this paper is to work with the full two-dimensional
systems, so that one is not restricted to radially-symmetric configurations.

One purpose of all this is to set up systems in which topological objects have
a high degree of stability, even when the lattice is quite coarse. So the
primary aim is not to simulate the continuum system, but rather to define
an alternative, ``genuine'' lattice system with, it is hoped, similar
properties (and more convenient to study numerically).

The two cases dealt with in this paper provide examples of two different kinds
of topological soliton. In the O(3) sigma-model, we have a ``texture'',
in which the field wraps the whole of two-dimensional space around the
target space $S^2$. By contrast, the vortex in the abelian Higgs model is
a ``monopole'', where the topology arises purely from the behaviour of the
field at spatial infinity. (But of course the Bogomol'nyi bound on its energy
involves the field throughout space.)

\bn{\bf2. The Lattice O(3) Sigma Model}

\sn Let us begin with a brief review of the continuum O(3) sigma model in
two space dimensions. The field $\vp$ is a unit 3-vector field on $\r2$
(\ie\ a smooth function from $\r2$ to the unit sphere $S^2$), with the
boundary condition $\vp\to\vp_0$ as $r\to\infty$ in $\r2$. Here $\vp_0$
is some given (fixed) point on the image sphere $S^2$. Any such field $\vp$ is
labelled by its winding number $k$, which represents the number of times
$\r2$ is wrapped around $S^2$. The energy of $\vp$ is
$$
   E_{{\rm cont}} = \int_{\r2} \half (\pa_j\vp)\cdot(\pa_j\vp)\,dx\,dy\,,
                      \eqno(2.1)
$$
and the appropriate Bogomol'nyi argument [3] gives the bound
$E_{{\rm cont}} \geq4\pi|k|$.

There are fields which attain this lower bound (such minimum-energy fields
will be called solitons in what follows). Since
$ E_{{\rm cont}}  $ is
invariant under the scaling transformation $\vp(x^j)\mapsto\vp(\lambda x^j)$,
these configurations are metastable rather than stable (their size is not
fixed); and this absence
of stabilty shows up when ones studies the dynamics of solitons [7]. One 
can modify the system in order to achieve stability: for example, by adding to
$ E_{{\rm cont}}$ a Skyrme term which prevents the soliton from becoming too
localized, and a potential term which prevents it from spreading out [15].
This, of course, raises the energy significantly above the Bogomol'nyi bound.

Another way of preventing the soliton from spreading out is to put it in a box;
in other words, we take the domain of $\vp$ to be a bounded region such as
$|x|\leq L$, $|y|\leq L$. The boundary condition is now $\vp=\vp_0$
around the edge of this region. The topological classification and the
Bogomol'nyi bound work just as before. This time, however, the bound cannot be
attained (except in the trivial case $k=0$), and all configurations are
unstable to shrinking. The addition to the energy of a term which opposes
shrinking (such as a Skyrme term) will ensure stability; the size of the
soliton is then determined by the balance between the expansionary effect
of this term and the containing effect of the walls of the region.

Let us now replace $\r2$ by a (bounded or unbounded) square lattice, where
the spatial variables $x$ and $y$ run over integer values. The field
$\vp(x,y)$ at the lattice site $(x,y)$ is a unit 3-vector, as before.
Instead of the partial derivative $\pa_x\vp$, we work with the inner product
$\vp\cdot\vp_x$, where $\vp_x$ denotes the nearest neighbour in the positive
$x$-direction [\ie\ $\vp_x(x,y)=\vp(x+1,y)$]; and similarly for $\pa_y\vp$.
Take the energy of $\vp$ to have the form (which couples only nearest-neighbour
spins)
$$
 E = \sum_{x,y}\bigl[ f(\vp\cdot\vp_x) +  f(\vp\cdot\vp_y)\bigr]\,, \eqno(2.2)
$$
where $f$ is a suitable function, to be described presently. The boundary
condition, as before, is $\vp=\vp_0$ around the rim of the lattice (which might
be at infinity).
The function $f:[-1,1]\to R$ is chosen to satisfy the following two conditions.
\item{(a)} {\it The energy should be positive-definite.} So we want $f(\xi)>0$
for $\xi<1$, and $f(1)=0$ (the trivial solution $\vp\equiv\vp_0$ then has zero
energy).
\item{(b)} {\it The lattice energy (2.2) reduces to (2.1) in the continuum
limit.} There is no scale in this system (that is why, without loss of
generality, the
lattice spacing was taken to be unity), and ``continuum limit'' means
$\vp\cdot\vp_x\to1$ and $\vp\cdot\vp_y\to1$ at all lattice sites. Using
$\half(\pa_x\vp)^2\approx1-\vp\cdot\vp_x$ reveals that we need $f'(1)=-1$ in
order to get the correct continuum limit.

\medskip

The most obvious function $f$ with these two properties is $f_{{\rm Heis}}(\xi)
=1-\xi$, which gives the well-known Heisenberg model. But in this case there
is no stable (or even metastable) configuration other than the trivial one
$\vp\equiv\vp_0$. Any ``topologically non-trivial'' configuration will
``unwind'', essentially because it is not costly enough for the directions
of neighbouring spins to be very different. In order to achieve topological
stability, one has to use a different $f$,
in particular making it more expensive for neighbouring spins
to be different. This aspect is discussed further below.

The immediate aim, though, is to have a lattice Bogomol'nyi bound; it turns
out that this also
ensures a measure of topological stability. So we add a further condition.
\item{(c)} {\it If the lattice configuration has winding number $k$, then
$E\geq4\pi|k|$.}

\ni For this to make sense, the winding number has to be well-defined, which
requires some restriction on the field $\vp$. The idea here is to say that
$\vp$ is {\it continuous} if the angle between nearest-neighbour spins is less
than $\pi/2$ (in other words, $\vp\cdot\vp_x>0$ and $\vp\cdot\vp_y>0$ at all
lattice sites). For a continuous $\vp$, the winding number $k$ is well-defined,
and corresponds to the number of times the lattice is wrapped around $S^2$; see
appendix A for more details.

A function $f$ which possesses all three properties (a), (b), (c) is
$$
   f(\xi) = \pi - 4\tan^{-1}\sqrt{\xi}\,; \eqno(2.3)
$$
this is proved in appendix A. Actually, (2.3) defines $f$ only for continuous
fields, where $\xi\in(0,1]$, these being our main concern here. One may obtain
an $f$ for the full range $\xi\in[-1,1]$, and therefore for all fields $\vp$,
by extending the above $f$ monotonically (for example, take the argument
of the arctan to be $-\sqrt{-\xi}$ if $\xi\leq0$).

Note that if one starts with a random configuration (for example, in a
lattice-statistical study), then this configuration is unlikely to be
continuous according to the definition given above. But when the field
is relaxed (cooled), then a ``steep'' function $f$ (such as the one
described above) will rapidly drive it towards continuity. This is certainly
not the case for the Heisenberg choice $f_{{\rm Heis}}$; in fact, the opposite
is true, as is illustrated in more detail below.

So if $\vp$ is a continuous field of spins on the lattice, with
winding number $k$, then its energy $E$ (defined by 2.2, 2.3) satisfies
$E\geq4\pi|k|$. Unlike in the continuum case, this lower bound cannot be
attained (see appendix A): the energy of any configuration with winding number
$k\neq0$ is strictly greater than $4\pi|k|$. On the infinite lattice $Z^2$,
such a configuration is unlikely to be stable. By spreading out in space,
it can approach (but never reach) the continuum limit, where the energy equals
$4\pi|k|$. In the continuum system, spreading only occurs if one puts in
something like a Skyrme term; but for this lattice system, no such extra term
is present. The function $f$ itself provides a spreading force, and a potential
barrier against topological decay (which is caused by shrinking).
The crucial feature is the behaviour of
$f(\xi)$ as $\xi\to0$ (in other words, as the field $\vp$ approaches a
discontinuity): the slope of $f(\xi)$ tends to $-\infty$. The corresponding
force drives the field away from the potential discontinuity and topological
decay. In the Heisenberg case, by contrast, the slope is $-1$, and this is
not enough to prevent discontinuity and decay: a continuous configuration
will evolve into a discontinuous one [8]. This can be
illustrated by looking at the energy of a one-parameter family of
configurations in which $\vp$ is
fixed at all lattice sites except the four around a particular plaquette.
Around this plaquette (cf.~figure 3), take
$$\eqalign{
           \vp(1) &= {1\over\sqrt2}(\nu, \nu, -2\eta) \cr
           \vp(2) &= {1\over\sqrt2}(-\nu, \nu, -2\eta) \cr
           \vp(3) &= {1\over\sqrt2}(\nu, -\nu, -2\eta) \cr
           \vp(4) &= {1\over\sqrt2}(-\nu, -\nu, -2\eta)\,, \cr
}$$
where $\eta\geq0$ is a small parameter and $\nu=1-\eta^2$.
Discontinuity occurs if $\eta=0$, for then $\vp(1)\cdot\vp(2)=0$, etc.
The boundary value of $\vp$ is
$\vp_0=(0,0,1)$, and $\vp\cdot\vp_0>0$ for all other lattice sites.
Then
$$
  {dE\over d\eta}\biggl|_{\eta=0} = 8\sqrt2 \lim_{\xi\to0}\bigl[
                                   \sqrt{\xi}f'(\xi)\bigr] + M\,,\eqno(2.4)
$$
where $M$ is a positive constant (depending on the values of $\vp$ at the
eight lattice sites linked to the four variable ones).
For stability, we want (2.4) to be negative, so that the field flows away
from the danger point $\eta=0$. In the Heisenberg case, the opposite will
clearly be the case, since $f'(0)$ is finite. One needs $f'(\xi)\to-\infty$ at
least as fast as $1/\sqrt{\xi}$, and the function (2.3) has this property.

With our choice of $f$, a soliton does not shrink; but
on an infinite lattice, it will spread out indefinitely. In order to
stabilize its size, we can (as in the continuum case) either confine it to a
lattice of bounded extent, or introduce a potential term into the expression
for the energy. More details of the latter possibility are given in [17].
In that paper, the choice $f(\xi)=-\log\xi$ was made. This function is
greater than (2.3), and so the Bogomol'nyi bound is still valid; in addition,
the energy tends to infinity as $\vp$ approaches discontinuity, so there is
a higher degree of stability.

In order to stay as close to the Bogomol'nyi bound as possible, let us use
the ``minimal'' choice (2.3), and avoid an extra term in the energy.
Rather, we restrict to a finite lattice $1\leq x\leq n$,  $1\leq y\leq n$.
If $n\geq6$, then there exist continuous configurations with winding number
$k=1$. Starting with such configurations, the energy $E$ was minimized
numerically in order to find the corresponding solitons,
for a range of values of $n$ (from 6 to 82). The energy $E(n)$ of the soliton
depends on $n$, and is greater than $4\pi$, with (as one would expect)
$E(n)\to4\pi$ as $n\to\infty$. More precisely, the numerical result is
$E(n)\approx4\pi(1+2.1/n+4/n^2)$ for $n>20$.

Since the field maps each lattice site to a point on the unit sphere $S^2$,
one way of visualizing it is to think of a square net made of
``elastic'' fibres, each with natural length zero, wrapped around $S^2$. The
boundary of the net is gathered together at the single point $\vp_0$.
Topological
decay occurs if a plaquette of the net becomes a hemisphere --- the net
then slips off and collapses to the single point $\vp_0$. If the fibres obeyed
Hooke's law, then the net would indeed slip off in this way: the restoring
force in the fibres would not be strong enough to prevent them stretching too
much. A ``Heisenberg'' net is even worse --- the restoring force is weaker.
The potential energy stored in a fibre of length $d$ is $f(\cos^{-1}d)$; 
with $f$ as in (2.3), this is larger than the
Hooke's law value of $\half d^2$, and large enough to stabilize the net.
The total energy $E$ is minimized (locally) by a non-trivial configuration,
and the corresponding net is depicted in figure 1, for $n=12$.
(The picture should, strictly speaking, have the vertices joined by segments
of great circle, rather than by straight lines as shown.)

\bn{\bf3. A Lattice Abelian Higgs System.}

\sn As in section 2, we begin with a brief review of the continuum case.
The fields consist of a U(1) gauge potential $A_j$, and a complex-valued
scalar field $\psi$, both smooth on $R^2$. The covariant derivative of $\psi$
is $D_j\psi = \pa_j\psi - iA_j\psi$, and the magnetic field strength is
$$
  B = \ve^{jk} \pa_j A_k = \pa_x A_y - \pa_y A_x.
$$
The energy of the field $(A_j,\psi)$ is
$$
  E_{{\rm cont}}
    = \int_{R^2} \bigl[ \half|D_x\psi|^2 + \half|D_y\psi|^2 + \half B^2
        + {\textstyle{1\over8}}(|\psi|^2-1)^2 \bigr]\,dx\,dy\,,\eqno(3.1)
$$
and the boundary conditions are $|\psi|\to1$, $D_j\psi\to0$ (which imply
$B\to0$) as $x,y\to\infty$ in $R^2$. All units have been fixed, and the
factor of ${1\over8}$ in the potential term means that we are restricting
to the case of critical coupling.

The boundary conditions lead to a topological classification as follows.
If $C$ is a large circle $r=r_0\gg1$ in $R^2$, then $|\psi|\approx1$ on $C$;
so $\psi\bigr|_C$ is effectively a mapping from $C$ to the unit circle in the
complex plane, and it has a winding number $k$. Secondly, the boundary
condition  $D_j\psi\to0$ implies (via Stokes's theorem) that the total
magnetic flux $\Phi = \int B\,dx\,dy$ satisfies
$$
   \Phi = 2\pi k. \eqno(3.2)
$$
Finally, the Bogomol'nyi argument [1] implies that $E\geq\half|\Phi|$.
Combining this with (3.2) gives $E\geq\pi|k|$.

This lower bound on $E$ is attained for static (multi-)vortex configurations.
The vortex is stable; in particular, its size is fixed, and is (a few times)
unity. So this system (unlike the sigma model)
has a scale, and the lattice version will consequently contain a parameter
$h$ denoting the lattice spacing.

The most natural way to set up the system on a lattice is to use
lattice gauge theory. The
variables $x$ and $y$ now run over integer multiples of $h$, and label the
lattice sites. The complex scalar field $\psi$ is defined at each lattice site.
We want the system to be invariant under gauge transformations
$\psi \mapsto \widehat\psi = \Lambda \psi$,
where $\Lambda$ is a function from the lattice to U(1) (a phase at each lattice
site). This is achieved in the standard way, in which a phase is associated
with each link of the lattice. So on the link joining $(x,y)$ to $(x+h,y)$,
we have a phase $U^x(x,y)\in{\rm U(1)}$; and similarly $U^y(x,y)$ lives on
the link from $(x,y)$ to $(x,y+h)$. The covariant derivative of $\psi$
is represented by the gauged forward difference
$$
  \eqalign{ D_x \psi &= h^{-1}(\psi_x - U^x\psi), \cr
            D_y \psi &= h^{-1}(\psi_y - U^y\psi), \cr }
$$
where subscripts denote forward shifts on the lattice:
$\psi_x(x,y)=\psi(x+h,y)$ etc. Given that $U^x$ transforms under a gauge
transformation as
$U^x\mapsto\widehat{U^x}=\Lambda_x\Lambda^{-1}U^x$,
the covariant derivative transforms as
$D_x \psi \mapsto \Lambda_x D_x \psi$.
The magnetic field $B$ is defined in terms of the gauge-invariant product
$$
  \exp(iB)=U^x U^y_x (U^x_y)^{-1} (U^y)^{-1} \eqno(3.3)
$$
of four phases around a plaquette (in an anticlockwise sense).

Many lattice gauge theory studies of vortices have been made. Usually, the
$\int\half B^2$ term in (3.1) is replaced by the Wilson action
$\sum h^{-2} (1-\cos B)$ on the lattice; the Bogomol'nyi bound is then no
longer valid (cf.~appendix B). For example, studies of vortex scattering
were carried out in this way [11, 12]. In those cases, the lattice parameter
$h$ was taken to be relatively small (typically $h=0.15$,
about ${1\over20}$ the size of a
vortex), since the authors were modelling the continuum system; consequently,
problems of vortex instability did not arise. By contrast, in cases where
$h$ is larger, vortices become unstable and can disappear: see, for example,
ref 4, where $h$ is taken to be $\sqrt2$ or 1.

In order to have a well-defined topological charge, we need the quantity $B$
defined by (3.3) to be unambiguous [9, 13, 14]. Accordingly,
we require that for each plaquette, the product of the $U$-factors around it
should not equal $-1$, and then take $B$ to lie in the range $(-\pi,\pi)$.
A gauge field with this property is said to be {\it continuous\/}
(cf. [9, 13, 14]). Note that in the continuum limit $h\to0$, we have $B\to0$,
provided that the continuum gauge potential is continuous in the usual sense.

Let us now derive the analogue of (3.2) for the lattice case. For convenience,
we use a
finite lattice: $x$ and $y$ range from $-L$ to $L$, where $L\gg1$ (bearing in
mind that the
size of a vortex is of order unity). The boundary condition on the
field is taken to be analogous to the continuum case, namely:
\item{$\bullet$} $|\psi|=1$ for all vertices on the boundary;
\item{$\bullet$} $D_x\psi=0$ for all $x$-links on the boundary, \ie\ if
                 $y=\pm L$;
\item{$\bullet$} $D_y\psi=0$ for all $y$-links on the boundary, \ie\ if
                 $x=\pm L$.
\sn Define the total magnetic flux $\Phi$ to be the sum of $B$ over all
plaquettes, namely $\Phi=\sum_{x,y}B$, where $x$ and $y$ range, in
integer multiples of $h$, from $-L$ to $L-h$. So $\exp(i\Phi)$ is a product
of $U$-factors, and it is clear that for any two adjacent plaquettes, the
$U$-factors associated with their common link cancel. Consequently,
$\exp(i\Phi)$ is a boundary expression
$$
  \exp(i\Phi) = \Bigl(\prod_{x=-L}^{L-h}U^x\bigl|_{y=-L}\Bigr)
      \Bigl(\prod_{y=-L}^{L-h}U^y\bigl|_{x=L}\Bigr)
       \Bigl(\prod_{x=-L}^{L-h}(U^x)^{-1}\bigl|_{y=L}\Bigr)
         \Bigl(\prod_{y=-L}^{L-h}(U^y)^{-1}\bigl|_{x=-L}\Bigr). \eqno(3.4)
$$
Now for an $x$-link on the boundary, we have $0=hD_x\psi=
\psi_x-U^x\psi$, and hence $U^x=\psi_x/\psi$; similarly for $U^y$.
So in fact the right-hand side of (3.4) equals 1, by virtue of the
single-valuedness of $\psi$. Therefore $\Phi = 2\pi k$ for some integer
$k$, which we {\it define}\/\ to be the winding number of the field.

If the values of $\psi$ at all pairs of
neighbouring lattice sites on the boundary are not antipodal (\ie\ 
$\psi_x\neq-\psi$ or $\psi_y\neq-\psi$), then the winding number of $\psi$
can be defined more directly, and is equal to $k$. Namely, one adds up the
angles $-i\log(\psi_x/\psi)$, as one traverses the boundary in an
anticlockwise direction.

The next task is to devise an expression for the energy $E$ of the lattice
system which reduces to (3.1) in the continuum limit, and which satisfies
the Bogomol'nyi bound $E\geq\half|\Phi|$. In appendix B, it is proved that
the following expression has those properties:
$$
  E = h^2\sum_{x,y=-L}^{L-h}\bigl[ \half|D_x\psi|^2 + \half|D_y\psi|^2
        +\half h^{-4} g(h) B^2 + {\textstyle{1\over8}} |\Psi^2-1|^2
                \bigr]\,, \eqno(3.5)
$$
where $g(h) = \sqrt{1+h^4/4}$ and
$\Psi^2 = U^x \overline{\psi_x} (U^y)^{-1} \psi_y$. One's first guess might
have been the expression (3.5) with $\Psi^2$ replaced by $|\psi|^2$; and with
$\sum\half h^{-2}g(h)B^2$ replaced by the Wilson action mentioned previously,
or by the Manton [10] action $\sum\half h^{-2}B^2$. Studies such as [4, 11, 12]
used the Wilson form of this first guess.
The given formula represents a slight modification of
that (a modification which vanishes in the continuum limit). Note that
$\Psi^2$ is gauge-invariant, as  indeed is $E$.

The conclusion, then, is that if the gauge-Higgs system satisfies the stated
boundary conditions, and the gauge field is continuous, then the energy $E$
is bounded below by $\pi|k|$, where the integer $k$ is the winding number of
the system. This statement remains true if $L\to\infty$, \ie\ for a
lattice of infinite extent. In practice, vortices are exponentially localized
with unit size, and so the parameter $L$ is effectively irrelevant as long as
$L\gg1$. The system depends only on the dimensionless parameter $h$, with
$h\to0$ being the continuum limit.

Minimum-energy configurations (vortices) with $k=1$ were found numerically,
for a number of values of $h$ in the range from 0.5 to 1.  This involved
minimizing the function $E$ (using a conjugate-gradient method), after fixing
the gauge --- the ``radial'' gauge $U^x=\exp(-iy\gamma)$,
$U^y=\exp(ix\gamma)$ was used, with $\gamma(x,y)$ being a real-valued
function on the lattice. The variables $x$ and $y$ were taken to range
over odd-integer multiples of $h/2$, with the vortex being ``located'' at
$x=y=0$ (which is at the centre of a plaquette).

The energy density, in the case $h=1$, is depicted in figure 2.  The
expression (3.5) for $E$ is a sum over plaquettes; the lattice function
plotted in figure 2 is the average, over the four plaquettes
containing the lattice site $(x,y)$, of the summand $e(x,y)$ of (3.5).
Presenting the picture in this way illustrates the fact that the vortex is
located near the centre of a particular plaquette (the Higgs field winds once
around that plaquette). Also apparent is an $(x,y)\mapsto(-x,-y)$
asymmetry: this arises because we are using forward (as opposed to backward)
differences on the lattice. One would get a more symmetric picture if one
replaced (3.5) by its average with the corresponding backward-difference
expression; the Bogomol'nyi bound would remain valid if one did so.

The energy $E$ of this lattice vortex depends on $h$. To within numerical
uncertainties, it has the quadratic form
$$
   E = \pi[ 1 + 0.070h^2 + 0.018h^4 ]
$$
for $0.5\leq h \leq1$. The Bogomol'nyi bound $E\geq\pi$ is not attained, but
of course $E(h)\to \pi$ in the continuum limit $h\to0$.

The space of {\it all\/} gauge-Higgs fields on the lattice is connected,
and so one does not have absolute topological stability. For example, one
could start with a $k=1$ vortex and deform it (so that the value of $B$ on the
central plaquette increased towards $2\pi$), ending up at
a configuration
which was gauge-equivalent to the trivial field, with zero energy.
Along this path in configuration space, the central $B$ would have to pass
through the disallowed value $\pi$. In fact, the space of {\it continuous\/}
fields (where $B$ never equals $\pi$) is disconnected, and its components are
labelled by the winding number $k$.

So the question of stability of the lattice vortex amounts to asking whether
$B$ gets driven towards a discontinuity. This will indeed happen
if $h$ becomes too large. In the $h=1$ example illustrated above, the
maximum value of $B$ (namely $B$ for the central plaquette) is about
$0.17\pi$. But as $h$ increases, the central plaquette grows to encompass
more and more of the vortex, and eventually $B$ reaches the forbidden value
$\pi$. By contrast, if $h$ is not too large, then it costs energy to increase
$B$ to a point of discontinuity, and the vortex is (relatively) stable.
The point to be emphasized is that the Bogomol'nyi lattice system
introduced above allows $h$ to be larger than the ``naive'' lattice version.

\bn{\bf4. Concluding Remarks}

\sn We have seen examples of ways of maintaining useful topological features
for systems on the two-dimensional lattice. It seems likely that one could do
something similar in three dimensions, and obtain lattice Bogomol'nyi versions
of (say) the Skyrme model and the Yang-Mills-Higgs (monopole) system.

If one is close to the continuum limit, in the sense that the lattice spacing
is small
compared to the size of the topological solitons, then there may not be much
difference between various lattice versions of the continuum system.
An advantage of the versions described in this paper is that the lattice
spacing can be relatively large, without compromising the stability of the
solitons. This should lead to different results in lattice-statistical studies
such as those of [4, 8].

In the context of classical soliton dynamics, one may study the interactions
of topological solitons on relatively coarse lattices, a computationally
simpler task than numerical simulation of
the corresponding continuum systems. There are
many different ways of introducing time dependence. If one uses equations
which are  second-order in time, and compares with relativistic dynamics
such as the vortex-scattering studies of [11, 12], then an important
difference is that moving solitons will radiate energy and gradually slow down.
In [11, 12] this effect was negligible, because the lattice spacing $h$ was
small. One expects it to increase with $h$, but the relevant time-scale
may still be large compared to that of the scattering process. For example,
this is the case in the topological lattice sine-Gordon system [16]. In
that system, however, there is no Peierls-Nabarro energy barrier, as
mentioned previously. By contrast, there will be such a barrier in the
two-dimensional systems of this paper, which will have some effect. Detailed
dynamical studies are needed to provide more precise answers.

\bn{\bf Appendix A.}

\sn This appendix describes how the winding number $k$ of a continuous
lattice O(3) field is defined, and establishes the Bogomol'nyi bound
$E\geq4\pi|k|$ on the energy (2.2, 2.3) of such a field.

The definition of the winding number is that of [2], namely as follows.
Let $p$ be a plaquette of the lattice, made up of two triangles $p_a$ and
$p_b$ (see fig 3). The image of $p_a$ is a spherical triangle on $S^2$;
let $A(p_a)$ denote the signed area of this spherical triangle. In other
words, $A(p_a)$ is positive if the sequence $\vp(1)$, $\vp(2)$, $\vp(3)$
goes round anticlockwise as depicted in fig 3, and negative if the
sequence is clockwise. Let  $A(p_b)$ be defined similarly. Then the winding
number $k$ is defined by adding up these signed areas and dividing by the area
of the sphere, \ie
$$
  k = {1\over4\pi} \sum_p \bigl[ A(p_a) + A(p_b) \bigr]\, .
$$
The covering of the planar lattice by triangles gives a $|k|$-fold covering
of $S^2$ by the corresponding spherical triangles; in other words, $k$ is an
integer, and $|k|$ is the number of times the lattice is wrapped around the
image sphere.

This definition of $k$ breaks down if there is a triangle ($p_a$, say) for
which the sign of $A(p_a)$ is ambiguous. This would happen if the three
points  $\vp(1)$, $\vp(2)$, $\vp(3)$ lay on a great circle (so that the
spherical triangle became a hemisphere); in [2] such ``exceptional
configurations'' are excluded. In this paper, a stronger exclusion is
used: we restrict to {\it continuous} fields $\vp$, meaning that for each link
$\langle1-2\rangle$,
the angle between  $\vp(1)$ and $\vp(2)$ is acute (so the segment
of great circle joining  $\vp(1)$ to $\vp(2)$ has length less than $\pi/2$).
For a continuous field, the ambiguity described above cannot occur, and so the
winding number $k$ of such a field is well-defined.

Let us now apply some spherical trigonometry to obtain a bound on $A(p_a)$,
namely
$$
  A(p_a) \leq  |A(p_a)| \leq \half f(\cos B) + \half f(\cos C), \eqno(A1)
$$
where $f(\xi) = \pi - 4\tan^{-1}\sqrt{\xi}$, and where $B$ and $C$
are the lengths of two of the sides of the spherical triangle. [More
specifically, $\cos B = \vp(1)\cdot\vp(2)$ and  $\cos C = \vp(1)\cdot\vp(3)$.]
Let $\alpha$, $\beta$ and $\gamma$ be the internal angles, as depicted in
fig 3; the area ${\cal A} :=  |A(p_a)|$ is given by
${\cal A} =  \alpha + \beta + \gamma - \pi$.

In order to establish (A1), we start with
$$\eqalign{
           \sin\half{\cal A} &= -\cos\half( \alpha + \beta + \gamma) \cr
            &= \sin\half\alpha\,\sin\half(\beta + \gamma)
               - \cos\half\alpha\,\cos\half(\beta + \gamma). \cr
}$$
Now use the identity $\tan\half(\beta + \gamma)=K\cot\half\alpha$, where
$$
  K = {\cos\half(B-C)\over\cos\half(B+C)} > 1;
$$
this leads to
$$
 \sin\half{\cal A} = {(K-1)\cos\half\alpha\over\sqrt{1+K^2\cot^2\half\alpha}}.
                   \eqno(A2)
$$
For fixed $K$, the function (A2) has a maximum when
$\cos^2\half\alpha = 1/(K+1)$; and its maximum value is
$(K-1)/(K+1)=\tan(\half B)\tan(\half C)$. The conclusion so far, therefore,
is that
$$
   {\cal A} \leq 2\sin^{-1}\bigl[\tan(\half B)\tan(\half C)\bigr].
\eqno(A3)$$
The final step uses the following lemma (proof later):

\ni{\it Lemma:} $2\sin^{-1}(xy)\leq\sin^{-1}(x^2)+\sin^{-1}(y^2)$
for all $x,y\in[0,1]$.

\ni So (A3) becomes
$$
   {\cal A} \leq \sin^{-1}(\tan^2\half B) + \sin^{-1}(\tan^2\half C);
$$
and the identity $2\sin^{-1}(\tan^2\half B)=\pi-4\tan^{-1}\sqrt{\cos B}$
completes the
proof of the inequality (A1). Summing (A1), and the corresponding inequality
for $p_b$, over all plaquettes, then gives the bound $E\geq4\pi k$.

\ni{\it Proof of Lemma.} The function $F(x,y)=\sin^{-1}(x^2)+
\sin^{-1}(y^2) - 2\sin^{-1}(xy)$ is continuous on the square $x,y\in[0,1]$.
The gradient of $F$ vanishes only if $x=y$, and $F$ is zero on this line.
So we need only check that $F$ is non-negative on the boundary of the square,
for then $\min F = 0$. Clearly $F(x,0)$ and $F(0,y)$ are non-negative.
And $F(x,1)=G(x)$ is non-negative as well, as is $F(1,y)=G(y)$:
the function $G(x)=\pi/2+\sin^{-1}(x^2)-2\sin^{-1}x$ has a negative slope,
and $G(1)=0$. \qquad QED

The above derivation of a Bogomol'nyi bound also reveals how the function $f$
arises: it is the smallest function (giving the lowest energy) for which the
argument works. To see this, suppose that for a particular plaquette $p$,
we have $\beta=\gamma=\half\alpha$ (and so $B=C$). Then the bound (A3) is
attained, and hence $|A(p_a)|=f(\cos B)$. So for the given choice of $f$,
the bound for a single plaquette is attained if the image of that plaquette
is a spherical square. Of course, the image sphere $S^2$ cannot be tiled
with spherical squares, and so the total Bogomol'nyi bound $E\geq4\pi k$ can
never be attained.

\bn{\bf Appendix B.}

\sn The aim in this appendix is to prove that the lattice abelian-Higgs
energy (3.5) satisfies a Bogomol'nyi bound. The proof is analogous to that
of the continuum case, and indeed reduces to it as $h\to0$. The starting
point is to consider a single plaquette, and to study the quantity
$$
 {\cal E} = \quar|D_x\psi|^2 + \quar|D_y\psi|^2 +
             \quar|(D_x\psi)_y|^2 + \quar|(D_y\psi)_x|^2 +
             2 h^{-4} \sin^2(\half B) + {\textstyle{1\over8}}|\Psi^2-1|^2\,,
                \eqno(B1)
$$
which involves the Higgs field $\psi$, $\psi_x$, $\psi_y$, $\psi_{xy}$
at the four vertices of the plaquette, and the gauge field $U^x$, $U^y$,
$U^x_y$, $U^y_x$ on the four links around the plaquette. Note that
$(D_x\psi)_y$ denotes the $y$-shifted version of $D_x\psi$, namely
$$
(D_x\psi)_y = h^{-1}(\psi_x-U^x\psi)_y =  h^{-1}(\psi_{xy}-U^x_y\psi_y).
$$
Clearly $\cal E$ is closely related to the summand in (3.5). Note, however,
that $2h^{-4}\sin^2(\half B)$ corresponds to the Wilson action. We now observe
that $\cal E$ can be rewritten as
$$\eqalign{
 {\cal E} &= \quar\bigl| (U^x)^{-1}D_x\psi + i(U^y)^{-1}D_y\psi \bigr|^2 +
                \quar\bigl| (D_x\psi)_y + i(D_y\psi)_x \bigr|^2 \cr
      &\qquad + \half\bigl| h^{-2}[\exp(iB)-1] + \half i(\Psi^2-1)\bigr|^2 \cr
      &\qquad + \quar ih^{-1}\bigl[ (Y_x-Y) - (X_y-X)\bigr]
                + \half h^{-2} \sin B \,,\cr} \eqno(B2)
$$
where
$$\eqalign{
      Y = \psi U^y \overline{D_y\psi} - \overline\psi(U^y)^{-1}D_y\psi\,,\cr
      X = \psi U^x \overline{D_x\psi} - \overline\psi(U^x)^{-1}D_x\psi\,,\cr
}$$
with $Y_x$ and $X_y$ being, respectively, the $x$-shifted and $y$-shifted
versions of these. To check that (B1) and (B2) are equal is straightforward
algebra.

The first three terms in (B2) are non-negative. The fourth, when summed over
all plaquettes, gives zero; for example,
$$
  \sum_{x=-L}^{L-h} (Y_x-Y) = Y\bigl|_{x=L} -  Y\bigl|_{x=-L} = 0,
$$
since $D_y\psi$ vanishes for $x=\pm L$. So we deduce that
$$
  h^2\sum_{x,y=-L}^{L-h} {\cal E} \geq \half \sum_{x,y=-L}^{L-h} \sin B \,.
           \eqno(B3)
$$
This ``first attempt'' at a Bogomol'nyi bound does not quite work: the
right-hand side is not topological. One should replace $\sin B$
by $B$, so as to get $\half\sum B = \half\Phi$; and one should replace
$2h^{-4}\sin^2(\half B)$ on the left-hand side
by something which achieves the inequality $h^2\sum{\cal E}_{\rm new}
\geq\half\sum B$. Let us use the modified Manton action
$\half h^{-4}g(h) B^2$ on the left-hand side, where $g(h)$ is some function.
The desired inequality will be true as long as
$$
 \Xi(B) := \half h^{-2}g(h)B^2 - 2h^{-2}\sin^2(\half B)
           - \half B + \half\sin B \geq 0 \,. \eqno(B4)
$$
Indeed, adding the summed version of (B4) to (B3) gives
$E\geq\half\Phi$, where $E$ is defined by (3.5).

It remains only to check (B4). The second derivative
$$\eqalign{
   \Xi''(B) &= h^{-2} g(h) - h^{-2}\cos B - \half\sin B \cr
            &= h^{-2} \bigl[ g(h) - \sqrt{1+h^4/4}\,\sin(B+B_0) \bigr] \cr
}$$
is non-negative for all $B$, provided that $g(h)=\sqrt{1+h^4/4}$. Since
$\Xi(0)=0=\Xi'(0)$, we
then conclude that (B4) holds for all $B$. The given function
$g$ is not the smallest for which (B4) is true. But certainly $g$ has to be
greater than 1, with $g=1+O(h^4)$ for small $h$.

\bn{\bf References.}
\sn

\item{[1]} Bogomol'nyi, E. B.: The stability of classical solutions.
           Sov. J. Nucl. Phys. {\bf24}, 449--454 (1976)
\item{[2]} Berg, B., L\"uscher, M.: Definition and statistical distributions
            of a topological number in the lattice O(3) $\sigma$-model.
            Nucl. Phys. B {\bf190}, 412--424 (1981)
\item{[3]} Belavin, A. A., Polyakov, A. M.: Metastable states of
            two-dimensional isotropic ferromagnets.
            JETP Lett. {\bf22}, 245--247 (1975)
\item{[4]} Grunewald, S., Ilgenfritz, E.-M., M\"uller-Preussker, M.:
             Lattice vortices in the two-dimensional abelian Higgs model.
             Zeit. f\"ur Physik C {\bf33}, 561--568 (1987)
\item{[5]} Leese, R.: Discrete Bogomolny equations for the nonlinear
             O(3) $\sigma$-model in (2+1) dimensions. Phys Rev D {\bf40},
             2004--2013 (1989)
\item{[6]} Leese, R. A.: The stability of semilocal vortices at critical
             coupling. Phys. Rev D {\bf46}, 4677--4684 (1992)
\item{[7]} Leese, R. A., Peyrard, M., Zakrzewski, W. J.: Soliton stability
             in the O(3) $\sigma$-model in (2+1) dimensions. Nonlinearity
             {\bf3}, 387--412 (1990)
\item{[8]} L\"uscher, M.: Does the topological susceptibility in lattice
            $\sigma$ models scale according to the perturbative renormalization
            group? Nucl. Phys. B {\bf200}, 61--70 (1982)
\item{[9]} L\"uscher, M.: Topology of lattice gauge fields. Commun. Math.
            Phys. {\bf85}, 39--48 (1982)
\item{[10]} Manton, N. S.: An alternative action for lattice gauge fields.
           Phys. Lett. B {\bf96}, 328--330 (1980)
\item{[11]} Moriarty, K. J. M., Myers, E., Rebbi, C.: Dynamical interactions
             of cosmic strings and flux vortices in superconductors.
             Phys. Lett. B {\bf207}, 411--418 (1988)
\item{[12]} Myers, E., Rebbi, C., Strilka, R.: Study of the interaction
             and scattering of vortices in the abelian Higgs (or
             Ginzburg-Landau) model. Phys. Rev. D {\bf45}, 1355--1364 (1992)
\item{[13]} Panagiotakopoulos, C.: Topology of 2D lattice gauge fields.
            Nucl. Phys. B {\bf251}, 61--76 (1985)
\item{[14]} Phillips, A.: Characteristic numbers of $U_1$-valued lattice gauge
           fields. Ann. Phys. {\bf161}, 399--422 (1985)
\item{[15]} Piette, B. M. A. G., Schroers, B. J., Zakrzewski, W. J.:
             Multisolitons in a two-dimensional Skyrme model.
             Zeit. f\"ur Physik C {\bf65}, 165--174 (1995)
\item{[16]} Speight, J. M., Ward, R. S.: Kink dynamics in a novel sine-Gordon
            system. Nonlinearity {\bf7}, 475--484 (1994)
\item{[17]} Ward, R. S.: Stable topological Skyrmions on the 2D lattice.
           Lett. Math. Phys. {\bf35}, 385--393 (1995)
\item{[18]} Zakrzewski, W. J.: A modified discrete sine-Gordon model.
           Nonlinearity {\bf8}, 517--540 (1995)

\bye